\documentclass[aps,prl,twocolumn,showpacs,superscriptaddress,groupedaddress]{revtex4}  
\usepackage{graphicx}  
\usepackage{dcolumn}   
\usepackage{bm}        
\usepackage{amssymb}   

\usepackage{amsmath}
\usepackage{multirow}
\usepackage{subfig}

\def\ttbar{$t\bar{t}$\ }
\def\SME{{\text{SME}}}

\def\Etmiss{{E\kern-0.6em\slash}_{T}}

\begin{document}

\hspace{5.2in} \mbox{FERMILAB-PUB-12-085-E}

\hspace{5.2in} \mbox{IUHET 565}

\title{Search for violation of Lorentz invariance in top quark pair production and decay}
\affiliation{LAFEX, Centro Brasileiro de Pesquisas F\'{i}sicas, Rio de Janeiro, Brazil}
\affiliation{Universidade do Estado do Rio de Janeiro, Rio de Janeiro, Brazil}
\affiliation{Universidade Federal do ABC, Santo Andr\'e, Brazil}
\affiliation{University of Science and Technology of China, Hefei, People's Republic of China}
\affiliation{Universidad de los Andes, Bogot\'a, Colombia}
\affiliation{Charles University, Faculty of Mathematics and Physics, Center for Particle Physics, Prague, Czech Republic}
\affiliation{Czech Technical University in Prague, Prague, Czech Republic}
\affiliation{Center for Particle Physics, Institute of Physics, Academy of Sciences of the Czech Republic, Prague, Czech Republic}
\affiliation{Universidad San Francisco de Quito, Quito, Ecuador}
\affiliation{LPC, Universit\'e Blaise Pascal, CNRS/IN2P3, Clermont, France}
\affiliation{LPSC, Universit\'e Joseph Fourier Grenoble 1, CNRS/IN2P3, Institut National Polytechnique de Grenoble, Grenoble, France}
\affiliation{CPPM, Aix-Marseille Universit\'e, CNRS/IN2P3, Marseille, France}
\affiliation{LAL, Universit\'e Paris-Sud, CNRS/IN2P3, Orsay, France}
\affiliation{LPNHE, Universit\'es Paris VI and VII, CNRS/IN2P3, Paris, France}
\affiliation{CEA, Irfu, SPP, Saclay, France}
\affiliation{IPHC, Universit\'e de Strasbourg, CNRS/IN2P3, Strasbourg, France}
\affiliation{IPNL, Universit\'e Lyon 1, CNRS/IN2P3, Villeurbanne, France and Universit\'e de Lyon, Lyon, France}
\affiliation{III. Physikalisches Institut A, RWTH Aachen University, Aachen, Germany}
\affiliation{Physikalisches Institut, Universit\"at Freiburg, Freiburg, Germany}
\affiliation{II. Physikalisches Institut, Georg-August-Universit\"at G\"ottingen, G\"ottingen, Germany}
\affiliation{Institut f\"ur Physik, Universit\"at Mainz, Mainz, Germany}
\affiliation{Ludwig-Maximilians-Universit\"at M\"unchen, M\"unchen, Germany}
\affiliation{Fachbereich Physik, Bergische Universit\"at Wuppertal, Wuppertal, Germany}
\affiliation{Panjab University, Chandigarh, India}
\affiliation{Delhi University, Delhi, India}
\affiliation{Tata Institute of Fundamental Research, Mumbai, India}
\affiliation{University College Dublin, Dublin, Ireland}
\affiliation{Korea Detector Laboratory, Korea University, Seoul, Korea}
\affiliation{CINVESTAV, Mexico City, Mexico}
\affiliation{Nikhef, Science Park, Amsterdam, the Netherlands}
\affiliation{Radboud University Nijmegen, Nijmegen, the Netherlands}
\affiliation{Joint Institute for Nuclear Research, Dubna, Russia}
\affiliation{Institute for Theoretical and Experimental Physics, Moscow, Russia}
\affiliation{Moscow State University, Moscow, Russia}
\affiliation{Institute for High Energy Physics, Protvino, Russia}
\affiliation{Petersburg Nuclear Physics Institute, St. Petersburg, Russia}
\affiliation{Instituci\'{o} Catalana de Recerca i Estudis Avan\c{c}ats (ICREA) and Institut de F\'{i}sica d'Altes Energies (IFAE), Barcelona, Spain}
\affiliation{Uppsala University, Uppsala, Sweden}
\affiliation{Lancaster University, Lancaster LA1 4YB, United Kingdom}
\affiliation{Imperial College London, London SW7 2AZ, United Kingdom}
\affiliation{The University of Manchester, Manchester M13 9PL, United Kingdom}
\affiliation{University of Arizona, Tucson, Arizona 85721, USA}
\affiliation{University of California Riverside, Riverside, California 92521, USA}
\affiliation{Florida State University, Tallahassee, Florida 32306, USA}
\affiliation{Fermi National Accelerator Laboratory, Batavia, Illinois 60510, USA}
\affiliation{University of Illinois at Chicago, Chicago, Illinois 60607, USA}
\affiliation{Northern Illinois University, DeKalb, Illinois 60115, USA}
\affiliation{Northwestern University, Evanston, Illinois 60208, USA}
\affiliation{Indiana University, Bloomington, Indiana 47405, USA}
\affiliation{Purdue University Calumet, Hammond, Indiana 46323, USA}
\affiliation{University of Notre Dame, Notre Dame, Indiana 46556, USA}
\affiliation{Iowa State University, Ames, Iowa 50011, USA}
\affiliation{University of Kansas, Lawrence, Kansas 66045, USA}
\affiliation{Kansas State University, Manhattan, Kansas 66506, USA}
\affiliation{Louisiana Tech University, Ruston, Louisiana 71272, USA}
\affiliation{Boston University, Boston, Massachusetts 02215, USA}
\affiliation{Northeastern University, Boston, Massachusetts 02115, USA}
\affiliation{University of Michigan, Ann Arbor, Michigan 48109, USA}
\affiliation{Michigan State University, East Lansing, Michigan 48824, USA}
\affiliation{University of Mississippi, University, Mississippi 38677, USA}
\affiliation{University of Nebraska, Lincoln, Nebraska 68588, USA}
\affiliation{Rutgers University, Piscataway, New Jersey 08855, USA}
\affiliation{Princeton University, Princeton, New Jersey 08544, USA}
\affiliation{State University of New York, Buffalo, New York 14260, USA}
\affiliation{Columbia University, New York, New York 10027, USA}
\affiliation{University of Rochester, Rochester, New York 14627, USA}
\affiliation{State University of New York, Stony Brook, New York 11794, USA}
\affiliation{Brookhaven National Laboratory, Upton, New York 11973, USA}
\affiliation{Langston University, Langston, Oklahoma 73050, USA}
\affiliation{University of Oklahoma, Norman, Oklahoma 73019, USA}
\affiliation{Oklahoma State University, Stillwater, Oklahoma 74078, USA}
\affiliation{Brown University, Providence, Rhode Island 02912, USA}
\affiliation{University of Texas, Arlington, Texas 76019, USA}
\affiliation{Southern Methodist University, Dallas, Texas 75275, USA}
\affiliation{Rice University, Houston, Texas 77005, USA}
\affiliation{University of Virginia, Charlottesville, Virginia 22901, USA}
\affiliation{University of Washington, Seattle, Washington 98195, USA}
\author{V.M.~Abazov} \affiliation{Joint Institute for Nuclear Research, Dubna, Russia}
\author{B.~Abbott} \affiliation{University of Oklahoma, Norman, Oklahoma 73019, USA}
\author{B.S.~Acharya} \affiliation{Tata Institute of Fundamental Research, Mumbai, India}
\author{M.~Adams} \affiliation{University of Illinois at Chicago, Chicago, Illinois 60607, USA}
\author{T.~Adams} \affiliation{Florida State University, Tallahassee, Florida 32306, USA}
\author{G.D.~Alexeev} \affiliation{Joint Institute for Nuclear Research, Dubna, Russia}
\author{G.~Alkhazov} \affiliation{Petersburg Nuclear Physics Institute, St. Petersburg, Russia}
\author{A.~Alton$^{a}$} \affiliation{University of Michigan, Ann Arbor, Michigan 48109, USA}
\author{G.~Alverson} \affiliation{Northeastern University, Boston, Massachusetts 02115, USA}
\author{M.~Aoki} \affiliation{Fermi National Accelerator Laboratory, Batavia, Illinois 60510, USA}
% ...
%
%
\collaboration{The D0 Collaboration\footnote{with visitors from
$^{a}$Augustana College, Sioux Falls, SD, USA,
$^{b}$The University of Liverpool, Liverpool, UK,
$^{c}$UPIITA-IPN, Mexico City, Mexico,
$^{d}$DESY, Hamburg, Germany,
,
$^{e}$SLAC, Menlo Park, CA, USA,
$^{f}$University College London, London, UK,
$^{g}$Centro de Investigacion en Computacion - IPN, Mexico City, Mexico,
$^{h}$ECFM, Universidad Autonoma de Sinaloa, Culiac\'an, Mexico
and
$^{i}$Universidade Estadual Paulista, S\~ao Paulo, Brazil.
$^{\ddag}$Deceased.
}} \noaffiliation
\vskip 0.25cm

\date{March 27, 2012}

\begin{abstract}
Using data collected with the D0 detector at the Fermilab Tevatron Collider, corresponding to 5.3 fb$^{-1}$ of integrated luminosity, we search for violation of Lorentz invariance by examining the \ttbar production cross section in lepton+jets final states. We quantify this violation using the standard-model extension framework, which predicts a dependence of the \ttbar production cross section on sidereal time as the orientation of the detector changes with the rotation of the Earth. Within this framework, we measure components of the matrices $(c_Q)_{\mu\nu 33}$ and $(c_U)_{\mu\nu 33}$ containing coefficients used to parametrize violation of Lorentz invariance in the top quark sector. Within uncertainties, these coefficients are found to be consistent with zero.
\end{abstract}

\maketitle


We investigate the possibility of Lorentz invariance violation (LIV) in the top quark ($t$) sector, using data collected with the D0 detector at the Fermilab Tevatron $p\bar{p}$ Collider corresponding to 5.3 fb$^{-1}$ of integrated luminosity collected between August 2002 and June 2009. We examine events in which a \ttbar pair is produced and decays into a final state including two light quarks ($\bar{q},q'$), two $b$ quarks ($b,\bar{b})$, and a lepton-neutrino pair ($\ell, \nu_{\ell}$) via the mode \ttbar$\rightarrow W^+ b W^- \bar{b} \rightarrow \ell \nu_{\ell} b \bar{q} q' \bar{b}$, where $\ell = e,\mu$. The standard-model extension (SME) framework \cite{bib:SME} provides an effective field theoretical treatment for violation of Lorentz and CPT symmetry in particle interactions by introducing Lorentz-violating terms to the Lagrangian density of the standard model (SM). As yet, there are no quantitative limits on violations of CPT or Lorentz invariance in the top quark sector \cite{bib:Alan_tables}. This parameter space is accessible only at high-energy particle colliders. Because top quarks decay before hadronizing, this study also offers the possibility of extending such investigation to what are essentially free quarks.

Strong limits have been set on the magnitude of LIV in gravitational and electromagnetic interactions, as well as in many particle sectors. Most constraints in the particle sectors are for matter involving quarks of the first generation. There are also sensitive limits on SME coefficients for the second generation, but only a few for the third generation \cite{bib:Alan_tables}. The latter include limits for the $b$ quark through $B$-meson oscillations \cite{bib:Bmeson}, and for $\tau$ neutrinos from neutrino oscillations \cite{bib:tauNu}. There is also a constraint for $\tau$ leptons deduced from theoretical grounds using astrophysical observations \cite{bib:tauLep}. Constraints on LIV have also been predicted for the SM Higgs sector, as derived from radiative corrections \cite{bib:Higgs}. Many of the limits on LIV coefficients in the quark, lepton, and gauge boson sectors are $\lesssim$10$^{-5}$. However, no constraints have yet been placed on LIV in the top quark sector. Because the SME represents a general phenomenological formalism, the LIV terms of the SME are not constrained to couple with the same strength to all particle species. We therefore consider separately only those SME terms that affect the top quark fields in \ttbar events.

While it has been shown that CPT violation implies violation of Lorentz invariance \cite{bib:Greenberg}, the contributions from CPT violating terms in the SME to the matrix element for \ttbar production and decay are suppressed. However, contributions from other Lorentz-violating terms can be significant \cite{bib:CPT10}. At leading order in LIV coefficients, the matrix element describing the production and decay of a \ttbar pair involves coefficients of the form $c_{\mu\nu}$, where $\mu$ and $\nu$ refer to space-time indices. Although at leading order CPT\nobreakdash-odd SME terms describing LIV in the top quark sector are not observable in \ttbar production or decay, this analysis is sensitive to several components of the CPT\nobreakdash-even $(c_Q)_{\mu\nu AB}$ and $(c_U)_{\mu\nu AB}$ terms, where $A,B = 3,3$ refer to the third quark generation. The $(c_Q)_{\mu\nu 33}$ are the SME coefficients coupling to the left-handed components of the third generation quark fields, and $(c_U)_{\mu\nu 33}$ are the SME coefficients coupling to the right-handed singlet top quark field. For brevity, we drop the generation subscripts since we are restricting the analysis to the terms that couple to the top quark fields. To compare our results with SME studies in other particle sectors \cite{bib:Alan_tables}, we also examine the linear combinations

\begin{equation}
\begin{array}{c}
c_{\mu\nu} = (c_Q)_{\mu\nu} + (c_U)_{\mu\nu}, \\
d_{\mu\nu} = (c_Q)_{\mu\nu} - (c_U)_{\mu\nu}.
\end{array}
\label{eq:linCombs}
\end{equation}

The matrix element for leading-order \ttbar production and decay, including leading-order contributions from SME terms, can be written as \cite{bib:CPT10}

\begin{equation}
\lvert\mathcal{M}\rvert^2_{\SME} = P F \bar{F}  + (\delta P) F \bar{F} + P (\delta F) \bar{F} + P F (\delta\bar{F}).
\label{eq:ME}
\end{equation}

\noindent The $P$ terms are functions of the parton momenta at the \ttbar production vertex, while the $F$ terms involve parton momenta at the decay vertices. The $PF\bar{F}$ term corresponds to the usual SM component, while the $\delta$-terms reflect the dependence on SME coefficients. This expression summarizes how the SME modifies the matrix element for \ttbar production and decay at leading order.

The $\delta$-terms contain contractions of $c_{\mu\nu}$ coefficients with tensors that are functions of the four-momenta of the particles in \ttbar production and decay. Due to the $V-A$ structure of the weak current, the right-handed coefficients, $(c_U)_{\mu\nu}$, couple only to the production ($\delta P$) terms, while the left-handed coefficients, $(c_Q)_{\mu\nu}$, couple to both production and decay ($\delta F$) terms. The matrices of $c_{\mu\nu}$ coefficients are symmetric and traceless. Within the SME, these coefficients are defined by convention in the canonical Sun-centered reference frame \cite{bib:Alan_tables}.

The kinematic component of the $\delta$-terms of Eq.\ (\ref{eq:ME}) can be evaluated in any coordinate system. A convenient reference frame is that of a coordinate system fixed to the measuring apparatus, and we therefore choose to evaluate such contractions in the D0 coordinate system. In this system, the momenta entering the calculation of Eq.\ (\ref{eq:ME}) are just the momenta of the particles measured in the detector, and, to calculate the matrix element, the coefficients $(c_U)_{\mu\nu}$ and $(c_Q)_{\mu\nu}$ must therefore be transformed from the Sun's reference system to the D0 coordinate system.

Since the Earth is rotating about its axis, the transformation of the coefficients $(c_U)_{\mu\nu}$ and $(c_Q)_{\mu\nu}$ from the Sun-centered frame to the laboratory frame introduces a time dependence. The relevant time scale is the sidereal day, which has a period of 23 hr 56 min 4.1 s (86,164.1 s). If any of the coefficients $(c_U)_{\mu\nu}$ or $(c_Q)_{\mu\nu}$ are non-zero in the Sun-centered frame, they can be detected through a periodic oscillation in the number of \ttbar events observed in the Earth-based detector as a function of sidereal time.


The data used for this analysis correspond to 5.3 fb$^{-1}$ of integrated luminosity collected with the D0 detector. The D0 detector \cite{bib:D0-descr} consists of several subdetectors designed for identification and reconstruction of the products of $p\bar{p}$ collisions. A silicon microstrip tracker and central fiber tracker surround the interaction region for pseudorapidities $|\eta| < 3$ and $|\eta| < 2.5$, respectively (where $\eta = -\ln[\tan(\theta/2)]$ is measured relative to the center of the detector, and $\theta$ is the polar angle with respect to the proton beam direction). These elements of the central tracking system are located within a 2 T superconducting solenoidal magnet, providing measurements for reconstructing event vertices and paths of charged particles. Particle energies are measured using a liquid argon and uranium calorimeter. Outside of the calorimetry, trajectories of muons are measured using three layers of tracking detectors and scintillation trigger counters, with 1.8 T iron toroidal magnets between the first two layers. Plastic scintillator arrays in front of the end-calorimeter cryostats provide measurements of luminosity.

We employ the same event selection as described in greater detail in Ref.\ \cite{bib:xsec-d0note}. Briefly, events are collected using a suite of triggers selecting events with a single lepton ($e$ or $\mu$) or a single lepton plus a jet. Candidate \ttbar events in the lepton+jets channels are then selected by requiring the presence of one isolated electron (or muon) candidate with transverse momentum $p_T > 20$~GeV and pseudorapidity $|\eta| < 1.1$ ($2$), and an imbalance in transverse energy of $\Etmiss > 20$~GeV ($25$~GeV). Events are divided into bins of jet multiplicity, and all jets are required to be reconstructed with $p_T > 20$~GeV and $|\eta| < 2.5$, with a leading jet of $p_{T}>40$~GeV. One of the jets is required to be tagged as a $b$-jet candidate through a neural-network-based (NN) algorithm \cite{bib:btagging}. The time of production of each \ttbar event is recorded with the event data, with an average accuracy of approximately $\pm$ 30 s. To follow the conventions utilized in other SME studies \cite{bib:Alan_tables}, we shift the origin of the time coordinate to correspond to the vernal equinox of the year 2000.


The SME predicts time dependent effects on the \ttbar cross section of the form

\begin{equation}
\sigma(t) \approx \sigma_{\text{ave}}\left[ 1 + f_\SME(t) \right],
\label{eq:sigObs-final}
\end{equation}

\noindent where $\sigma_{\text{ave}}$ is the observed (time averaged) cross section for
\ttbar$\rightarrow W^+ b W^- \bar{b} \rightarrow \ell \nu_{\ell} b \bar{q} q' \bar{b}$,
in $\ell$+jets final states. To arrive at Eq.\ (\ref{eq:sigObs-final}), we compare the contribution from the SME terms in Eq.\ (\ref{eq:ME}) to the SM expectation by considering the ratio of $\lvert\mathcal{M}\rvert_\SME^2$ to the SM component $PF\bar{F}$. The SME contributions in this ratio are collected into the function

\begin{align}
f_\SME(t) = &\;\left[(c_Q)_{\mu\nu}+(c_U)_{\mu\nu}\right] R_\alpha^\mu(t) R_\beta^\nu(t) A_P^{\alpha\beta} \nonumber\\
         &\;+ (c_Q)_{\mu\nu} R_\alpha^\mu(t) R_\beta^\nu(t) A_F^{\alpha\beta}.
\label{eq:fSME-full}
\end{align}

\noindent Eq.\ (\ref{eq:fSME-full}) is a product of the matrices of time-independent coefficients $(c_Q)_{\mu\nu}$ and $(c_U)_{\mu\nu}$, four-by-four matrices of terms that depend on the event production ($A_P^{\alpha\beta}$) and decay ($A_F^{\alpha\beta}$) kinematics in the D0 frame, and a rotation matrix $R_\alpha^\mu(t)$ that transforms $A_P^{\alpha\beta}$ and $A_F^{\alpha\beta}$ from the D0 frame to the Sun-centered frame.

The $A_P^{\alpha\beta}$ and $A_F^{\alpha\beta}$ matrices are evaluated using \ttbar Monte Carlo events generated with {\sc pythia} \cite{bib:PYTHIA}. Events that pass detector acceptance, trigger, event reconstruction, and analysis selections (modeled by a full simulation of the D0 detector) are corrected according to the SME expectation of Eq.\ (\ref{eq:ME}).


The SME contribution to the cross section has the general form $f_\SME(t) = C_{\mu\nu} R_\alpha^\mu(t) R_\beta^\nu(t) A^{\alpha\beta}$ for the four model assumptions summarized in Table \ref{tab:fSME-cases}. For each model, $C_{\mu\nu}$ represents the constant coefficients we wish to determine, and $A^{\alpha\beta}$ refers to the appropriate linear combination of $A_P^{\alpha\beta}$ and $A_F^{\alpha\beta}$.

\begin{table}[ht]
  \begin{center}
    \caption{$f_\SME(t)$ for different SME assumptions.}\label{tab:fSME-cases}
    \begin{tabular}{cl}
      \hline
      \hline
      Assumption & \multicolumn{1}{c}{$f_\SME(t)$} \\
      \hline
      $(c_U)_{\mu\nu} = 0$ & $(c_Q)_{\mu\nu} R_\alpha^\mu(t) R_\beta^\nu(t)
                           (A_P^{\alpha\beta} + A_F^{\alpha\beta})$ \\
      $(c_Q)_{\mu\nu} = 0$ & $(c_U)_{\mu\nu} R_\alpha^\mu(t) R_\beta^\nu(t)(A_P^{\alpha\beta})$ \\
      $c_{\mu\nu} = 0$ & $d_{\mu\nu} R_\alpha^\mu(t) R_\beta^\nu(t) \frac{1}{2} A_F^{\alpha\beta}$ \\
      $d_{\mu\nu} = 0$ & $c_{\mu\nu} R_\alpha^\mu(t) R_\beta^\nu(t)
                        (A_P^{\alpha\beta} + \frac{1}{2} A_F^{\alpha\beta})$ \\
      \hline
      \hline
    \end{tabular}
  \end{center}  
\end{table}

For each model, we estimate one possible component of $C_{\mu\nu}$ at a time. We impose the requirements that each tensor $C_{\mu\nu}$ is symmetric and traceless, choosing $C_{XX} = -C_{YY}$ to satisfy the latter condition. We adopt the index ordering conventions $\mu,\nu=\{T,X,Y,Z\}$ to refer to coordinates in the Sun-centered frame and $\alpha,\beta=\{t,x,y,z\}$ for coordinates in the D0 frame. Evaluating Eq.\ (\ref{eq:fSME-full}) for the different assumptions of Table \ref{tab:fSME-cases} yields the following results: (i) Coefficients $C_{TT}$ and $C_{ZZ}$ contribute only to the total cross section, and we do not attempt to extract these coefficients. (ii) Coefficients $C_{TX}$, $C_{TY}$, and $C_{TZ}$ combine with the small off-diagonal elements of matrices $A_P^{\alpha\beta}$ and $A_F^{\alpha\beta}$, for which we expect poor sensitivity. (iii) Coefficients $C_{XZ}$ and $C_{YZ}$ couple to expressions that depend on sidereal time (differing by a phase of $\pi$/2). (iv) Coefficients $C_{XX}$ and $C_{XY}$ couple to time dependent expressions with twice the sidereal frequency, and the two terms differ by a phase of $\pi$/4.

Table \ref{tab:fSME-forms} collects the resulting forms of the function $f_\SME(t)$ for different assumptions. We refer to the ``sidereal phase'' $\omega_{s}t$ as $\phi$, where $\omega_{s}$ is the inverse of the sidereal day. The $b$-terms in these expressions depend on the colatitude of the detector, the orientation of the proton beam at the detector relative to geographic north, and the ${XX}$ and ${ZZ}$ elements of the combination of $A_P^{\alpha\beta}$ and $A_F^{\alpha\beta}$ that are appropriate to the particular assumption of the model.

\begin{table}[ht]
  \begin{center}
    \caption{Forms for $f_\SME(\phi)$ used to extract SME coefficients.}\label{tab:fSME-forms}
    \begin{tabular}{ll}
      \hline
      \hline
      Condition & $f_\SME(\phi)$ \\
      \hline
      $C_{XX}$ = $-C_{YY}$ & $2 C_{XX} \left( \frac{b_1-b_2}{2} \cos 2\phi + b_3 \sin 2 \phi \right)$ \\ [1ex]
      $C_{XY}$ =  $C_{YX}$ & $2 C_{XY} \left( \frac{b_1-b_2}{2} \sin 2\phi - b_3 \cos 2 \phi \right)$ \\ [1ex]
      $C_{XZ}$ =  $C_{ZX}$ & $2 C_{XZ} \left( b_4 \cos \phi + b_5 \sin \phi \right)$ \\ [1ex]
      $C_{YZ}$ =  $C_{ZY}$ & $2 C_{YZ} \left( b_4 \sin \phi - b_5 \cos \phi \right)$ \\ [1ex]
      \hline
      \hline
    \end{tabular}
  \end{center}
\end{table}


Assuming that any LIV originates from just the top quark sector, we expect the background rate (principally $W$+jets events) to be proportional only to the luminosity. To search for a signal varying with sidereal time, we sum the contributions to each of twelve $N_i$ bins (corresponding to two sidereal hours each) for all data:

\begin{equation}
N_i \approx N_{\text{tot}} \frac{\mathcal{L}_i}{\mathcal{L}_{\text{int}}} \left[ 1 + f_S f_\SME(\phi_i) \right],
\label{eq:Ni}
\end{equation}

\noindent where $N_{\text{tot}}$ is the total number of signal ($t\bar{t}$) and background (non-$t\bar{t}$) events corresponding to the total integrated luminosity $\mathcal{L}_{\text{int}}$, $\mathcal{L}_i$ is the integrated luminosity over the appropriate bin of sidereal phase $\phi_i$, and $f_S$ is the average fraction of signal events in the data.

We extract $f_S$ from the data that was used previously to determine the \ttbar cross section in $\ell$+jets events \cite{bib:xsec-d0note}. The \ttbar cross section is measured in bins of jet multiplicity for the $e$+jets and $\mu$+jets channels. The subset of events with at least four reconstructed jets that pass selection requirements contain a high fraction of \ttbar events, providing the best sensitivity to any time dependence in the \ttbar event rate. We find $f_S\text{($e$+$>$3-jets)} = 0.78 \pm 0.12$ and $f_S\text{($\mu$+$>$3-jets)} = 0.76 \pm 0.11$. Because of this difference, we treat the electron and muon channels separately.

To simplify fitting $f_\SME(\phi)$ to the data, we define a variable $R$ for each bin:

\begin{equation}
R_i \equiv \frac{1}{f_S}
           \left( \frac{ N_i / N_{\text{tot}} }{\mathcal{L}_i / \mathcal{L}_{\text{int}}} - 1 \right).
\label{eq:R}
\end{equation}

\noindent Equation (\ref{eq:R}) is the luminosity-corrected sidereally-binned relative \ttbar event rate, which can be compared directly to $f_{\SME}(\phi)$. In the absence of any significant sidereal time dependence, all the $R_i$ values should be consistent with zero, while a sidereal time dependence would produce a sinusoidal variation in this rate. The amplitude for any sinusoidal dependence is given by the product of an SME coefficient and a mixture of contributions from the rotation matrix and appropriate combination of elements from $A_P^{\alpha\beta}$ and $A_F^{\alpha\beta}$. This latter mixture also fixes the phase of the sinusoidal function in a fit to the data.

\begin{figure}[ht]
  \begin{center}
    \includegraphics[height=0.45\columnwidth]{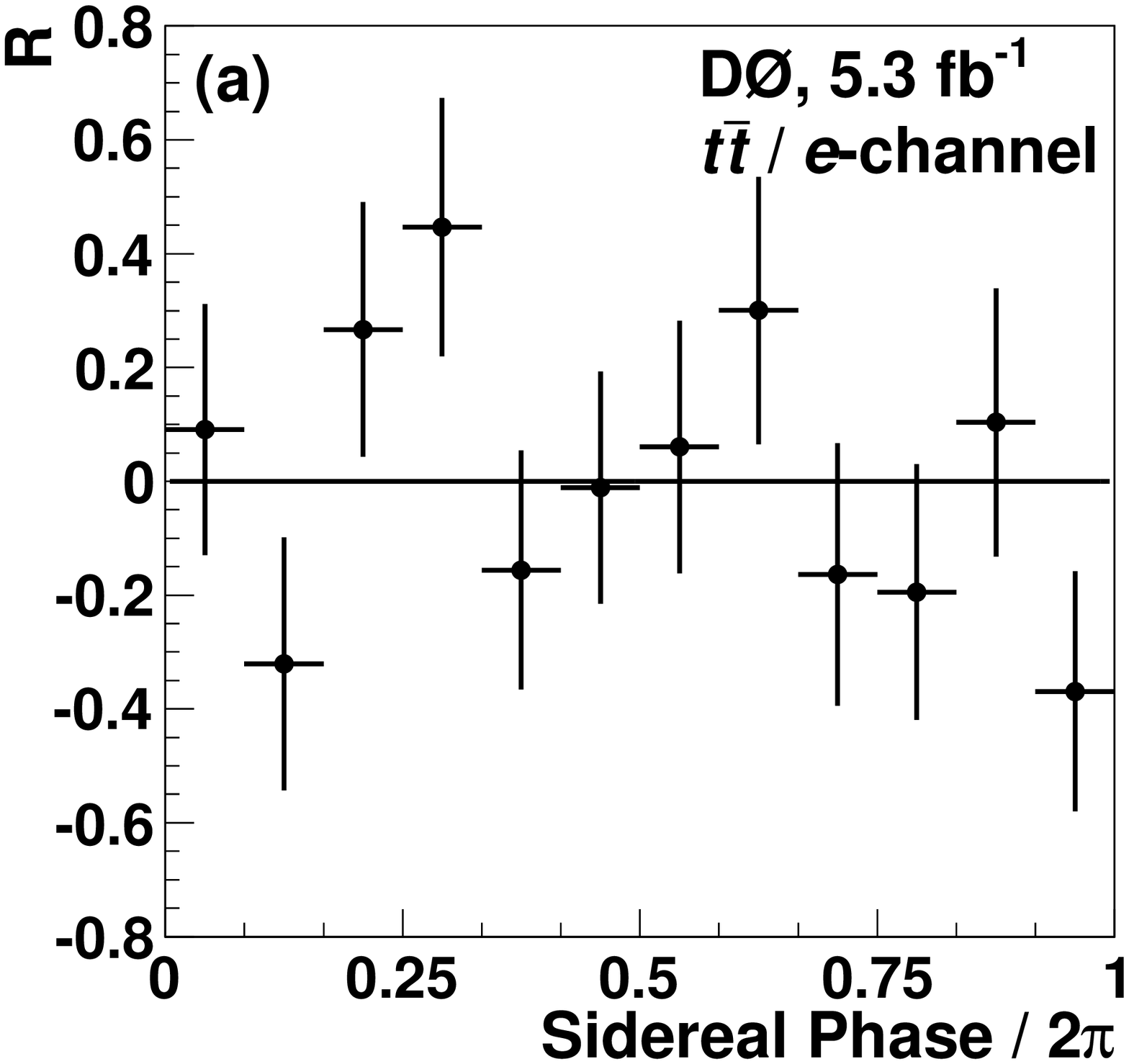}
    \includegraphics[height=0.45\columnwidth]{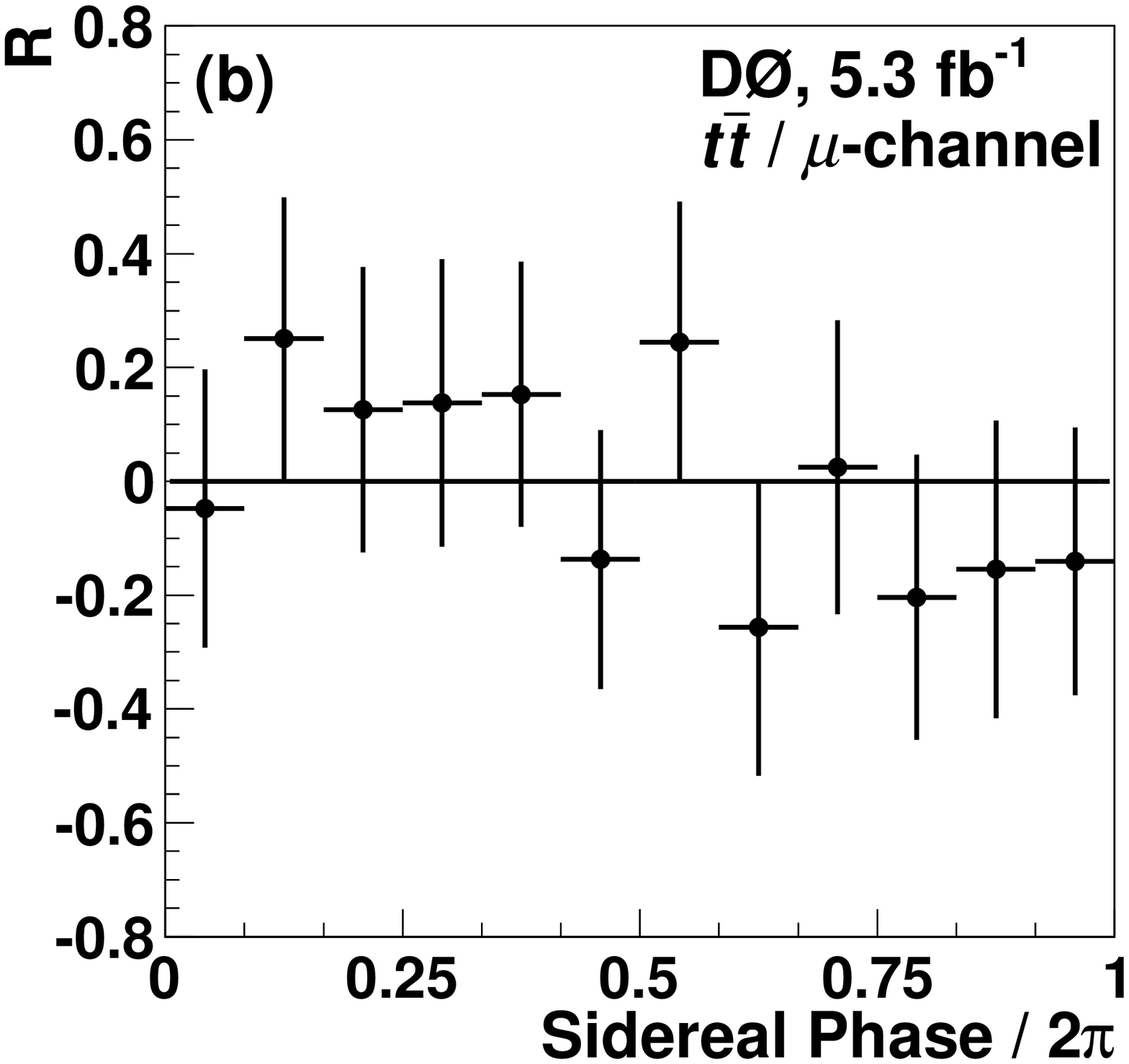}
  \end{center}
  \caption{The dependence of $R$, as defined in Eq.\ (\ref{eq:R}), on sidereal phase for
    (a) $e$+$>$3-jets \ttbar candidates, and (b) $\mu$+$>$3-jets \ttbar candidates.}
  \label{fig:datHists4}
\end{figure}

The resulting distributions for $R$ as a function of sidereal phase are shown in Fig.\ \ref{fig:datHists4}, separately for the electron and muon channels. The forms of $f_\SME(\phi)$ are fitted to these two distributions to estimate the values of the SME coefficients for the assumptions summarized in Tables \ref{tab:fSME-cases} and \ref{tab:fSME-forms}. We apply a small correction of 1.2\%--4.7\% to each extracted value to account for biases introduced by the finite bin size.

While the dominant contribution to the uncertainty on the SME coefficients results from the limited size of our \ttbar data sample, the estimated fraction of \ttbar events in the data contributes an additional uncertainty. We treat this as a systematic uncertainty in this study. The background from single top quark events can, in principle, exhibit SME effects. However, their relative contribution to the \ttbar sample is negligible ($\approx$1\%). The orientation and location of the detector, as well as the origin chosen for the time of events, also carry negligible uncertainties. Finally, any uncertainties in the values of the elements of $A_P^{\alpha\beta}$ and $A_F^{\alpha\beta}$ can potentially contribute a systematic uncertainty in this analysis, in ways similar to those discussed in the analysis of the \ttbar cross section, as summarized below.

The leading sources of systematic uncertainty in the kinematics of \ttbar events arise from: (i) the jet energy scale, (ii) jet energy resolution, and (iii) jet identification. These can affect the distributions of momenta reconstructed in the detector, but, as the elements of $A_P^{\alpha\beta}$ and $A_F^{\alpha\beta}$ reflect only average values of the components of the momenta over the detector acceptance, such averages are not very sensitive to small changes in kinematic parameters. The relative uncertainty of the contributing elements is negligible compared to the statistical uncertainty of the data and the systematic uncertainties on signal fractions $f_S$.

A periodic time dependence could potentially be introduced to the event rate through changes in event selection efficiency. We check this possibility by examining the luminosity-corrected sidereally-binned relative event rates ($R$ distributions) for the lepton+$n$-jets channels, where $n=2,3$. These bins of jet multiplicity contain relatively small contributions from \ttbar events, with $f_S\text{($\ell$+2-jets)} \approx 12\%$ and $f_S\text{($\ell$+3-jets)} \approx 45\%$. We extract the amplitudes for any time dependent oscillations, corresponding to the parameterizations used for the coefficients in Tables \ref{tab:Limits-cL}--\ref{tab:Limits-d}, in each of the four cross-check channels ($\ell$+$n$-jets, where $\ell=e,\mu$ and $n=2,3$). For each assumption, the ensemble of fits is consistent with no time dependence at levels of probability in the range 6\%--38\%. We therefore conclude that these cross checks give no indication of a sidereal time-dependent efficiency.

Finally, it should be noted that any residual non-sidereal time dependence is suppressed greatly by folding the data into twelve bins of sidereal phase, as the magnitude of any residual contribution following this folding depends inversely on the difference in the period of the time-dependent efficiency and the sidereal period. Most problematic would be an unexpected time-dependent efficiency with a period close to that of a sidereal day. The worst realistic case would be a contribution to detector efficiency that has a period of 24 solar hours. However, because the data taking spans approximately seven years, any contributions from such an effect would be suppressed by about a factor of 10. To affect our conclusions, we would have had to experience a highly unlikely periodic dependence of the efficiency of approximately $75\%$ over 24 hours. No periodic effects of this magnitude have ever been observed in the detection efficiencies for objects considered in this analysis.


Because the SME contribution to the matrix element is independent of lepton flavor, we perform a simultaneous fit to both the $e$+$>$3-jets and $\mu$+$>$3-jets data to obtain the final results. The extracted SME coefficients are all consistent with no time dependence, and we therefore find no evidence for violation of Lorentz invariance in the \ttbar system.

We define the observed limits (95\% C.L.\ intervals) for each SME coefficient as the extracted value $\pm$2 standard deviations. Because the magnitude of the 95\% confidence bounds on elements of the linear combination $c_{\mu\nu} = (c_Q)_{\mu\nu} + (c_U)_{\mu\nu}$ for the assumption of $d_{\mu\nu}=0$ are larger than 1, we cannot place meaningful limits on these combinations of SME coefficients in this analysis. The remaining limits are presented in Tables \ref{tab:Limits-cL}--\ref{tab:Limits-d}.

\begin{table}[ht]
  \begin{center}
    \caption{Limits on SME coefficients at the 95\% C.L.,
      assuming $(c_U)_{\mu\nu} \equiv 0$.}\label{tab:Limits-cL}
    \begin{tabular}{c r@{$\,\pm\,$}l@{$\,\pm\,$}l c}
      \hline
      \hline
      Coefficient & Value & Stat.\ & Sys.\ & 95\% C.L.\ Interval \\
      \hline
      $(c_Q)_{XX33}$ & $-0.12$ & $0.11$ & $0.02$ & $[-0.34,+0.11]$ \\
      $(c_Q)_{YY33}$ & $ 0.12$ & $0.11$ & $0.02$ & $[-0.11,+0.34]$ \\
      $(c_Q)_{XY33}$ & $-0.04$ & $0.11$ & $0.01$ & $[-0.26,+0.18]$ \\
      $(c_Q)_{XZ33}$ & $ 0.15$ & $0.08$ & $0.02$ & $[-0.01,+0.31]$ \\
      $(c_Q)_{YZ33}$ & $-0.03$ & $0.08$ & $0.01$ & $[-0.19,+0.12]$ \\
      \hline
      \hline
    \end{tabular}

    \caption{Limits on SME coefficients at the 95\% C.L.,
      assuming $(c_Q)_{\mu\nu} \equiv 0$.}\label{tab:Limits-cR}
    \begin{tabular}{c r@{$\,\pm\,$}l@{$\,\pm\,$}l c}
      \hline
      \hline
      Coefficient & Value & Stat.\ & Sys.\ & 95\% C.L.\ Interval \\
      \hline
      $(c_U)_{XX33}$ & $ 0.10$ & $0.09$ & $0.02$ & $[-0.08,+0.27]$ \\
      $(c_U)_{YY33}$ & $-0.10$ & $0.09$ & $0.02$ & $[-0.27,+0.08]$ \\
      $(c_U)_{XY33}$ & $ 0.04$ & $0.09$ & $0.01$ & $[-0.14,+0.22]$ \\
      $(c_U)_{XZ33}$ & $-0.14$ & $0.07$ & $0.02$ & $[-0.28,+0.01]$ \\
      $(c_U)_{YZ33}$ & $ 0.01$ & $0.07$ & $<0.01$ & $[-0.13,+0.14]$ \\
      \hline
      \hline
    \end{tabular}

    \caption{Limits on SME coefficients at the 95\% C.L.,
      assuming $c_{\mu\nu} \equiv 0$.}\label{tab:Limits-d}
    \begin{tabular}{c r@{$\,\pm\,$}l@{$\,\pm\,$}l c}
      \hline
      \hline
      Coefficient & Value & Stat.\ & Sys.\ & 95\% C.L.\ Interval \\
      \hline
      $d_{XX}$ & $-0.11$ & $0.10$ & $0.02$ & $[-0.31,+0.09]$ \\
      $d_{YY}$ & $ 0.11$ & $0.10$ & $0.02$ & $[-0.09,+0.31]$ \\
      $d_{XY}$ & $-0.04$ & $0.10$ & $0.01$ & $[-0.24,+0.16]$ \\
      $d_{XZ}$ & $ 0.14$ & $0.07$ & $0.02$ & $[-0.01,+0.29]$ \\
      $d_{YZ}$ & $-0.02$ & $0.07$ & $<0.01$ & $[-0.16,+0.13]$ \\
      \hline
      \hline
    \end{tabular}

  \end{center}
\end{table}

In the SME, different particles can have distinct Lorentz-violating properties, so it is of interest to test all species. Most constraints on LIV are for particles of the first and second generations, with a few limits on SME coefficients for the third generation. The only sector for which no constraints on Lorentz violation exist to date is the top quark \cite{bib:Alan_tables}. The limits on the $(c_Q)_{\mu\nu 33}$ and $(c_U)_{\mu\nu 33}$ coefficients determined in this study represent the first constraints on LIV in the top quark sector, and the first such constraints on any free quark.

%
We thank the staffs at Fermilab and collaborating institutions,
and acknowledge support from the
DOE and NSF (USA);
CEA and CNRS/IN2P3 (France);
MON, Rosatom and RFBR (Russia);
CNPq, FAPERJ, FAPESP and FUNDUNESP (Brazil);
DAE and DST (India);
Colciencias (Colombia);
CONACyT (Mexico);
NRF (Korea);
FOM (The Netherlands);
STFC and the Royal Society (United Kingdom);
MSMT and GACR (Czech Republic);
BMBF and DFG (Germany);
SFI (Ireland);
The Swedish Research Council (Sweden);
and
CAS and CNSF (China).
%
We also acknowledge support from the Indiana University Center for Spacetime Symmetries (IUCSS).


\end{document}